\documentclass[12pt,a4paper]{article}
\usepackage{graphicx}
\usepackage{cite}
\usepackage{a4wide}

\makeatletter

\renewcommand\theenumi{(\@roman\c@enumi)}

\makeatother

\begin{document}

\title{\textbf{FORMATION OF SOLITONS IN ATOMIC BOSE-EINSTEIN CONDENSATES
BY DARK-STATE ADIABATIC PASSAGE}}

\author{G.~Juzeli\=unas$^a$, J.~Ruseckas$^a$, P.~\"Ohberg$^b$,
M.~Fleischhauer$^c$\\
\textit{\small $^a$ Institute of Theoretical Physics and Astronomy of Vilnius
University,}\\
\textit{\small A.~Go\v{s}tauto 12, 01108 Vilnius, Lithuania}\\
\textit{\small $^b$ Department of Physics, School of Engineering and Physical
Sciences,}\\
\textit{\small Heriot-Watt University, Edinburgh EH14 4AS, UK}\\
\textit{\small $^c$  Fachbereich Physik, Technische Universit\"at
Kaiserslautern,}\\
\textit{\small D-67663 Kaiserslautern, Germany}}

\maketitle
\begin{abstract}
We propose a new method of creating solitons in elongated Bose-Einstein
Condensates (BECs) by sweeping three laser beams through the BEC.  If one of
the beams is in the first order (TEM10) Hermite-Gaussian mode, its amplitude
has a transversal $\pi$ phase slip which can be transferred to the atoms
creating a soliton. Using this method it is possible to circumvent the
restriction set by the diffraction limit inherent to conventional methods such
as phase imprinting.  The method allows one to create multicomponent (vector)
solitons of the dark-bright form as well as the dark-dark combination. In
addition it is possible to create in a controllable way two or more dark
solitons with very small  velocity and close to each other for studying their
collisional properties.

\vskip \baselineskip


\end{abstract}

\section{Introduction}

Atomic Bose-Einstein condensates (BECs) have received a great deal of interest
since they were first produced a decade ago
\cite{anderson95,bradley95,davis95}. They can exhibit various topological
excitations, such as vortices and solitons. The dynamics of solitons in
elongated BECs  \cite{stringari_book} is the atom-optics version of the
nonlinear propagation of light pulses in optical fibres \cite{kivshar98}.  The
BEC offers a remarkable freedom in terms of controlling the physical parameters
such as dimensionality and even the sign of the strength of the atom-atom
interaction \cite{stringari_book}.

Solitons in BECs can be of both dark and bright type. Dark solitons are formed
in BECs with repulsive interaction between the atoms \cite{stringari_book}. For
completely dark solitons the condensate wavefunction is zero at the centre and
changes its sign then crossing the central point, i.e.\ the condensate
wave-function has an infinitely steep $\pi$ phase slip at the centre
\cite{stringari_book}. On the other hand, the bright solitons are formed in
BECs with repulsive interaction between the atoms. The wave-function of the BEC
is then localised at the centre \cite{stringari_book} and goes to zero further
away from this point.  The dark solitons which  manifests themselves as  a
density minimum moving with a constant speed against a uniform background
density, as well as bright solitons which are shape preserving wave packets,
have both been experimentally realised
\cite{burger99,denschlag00,khaykovich02,strecker02}.  The dynamics of solitons
in BECs has been extensively studied.  This has included investigations of the
stability properties \cite{muryshev99}, as well as soliton dynamics in
inhomogeneous clouds \cite{busch00}, in multicomponent BECs
\cite{ohberg01,busch01} and in supersonic flow \cite{el06}.  Solitons can be
created in various ways with a variable degree of controllability, e.g., by
colliding clouds of BEC \cite{reinhardt97,scott98,brazhnyi03} or engineering
the density \cite{dutton01,ginsberg05}.

Traditionally dark solitons in BECs are created using phase imprinting
\cite{burger99,denschlag00,dobrek99,carr01,biao02}, where a part of the
condensate cloud is illuminated by a far detuned laser pulse in order to induce
a sharp $\pi$ phase slip in the wave function. The subsequent dynamics can
indeed develop solitons \cite{burger99,denschlag00}.  There are, however, some
rather severe drawbacks with such a method of phase engineering. The resolution
of the required phase slip is naturally restricted by the diffraction limit,
i.e.\ the width of the phase slip should be larger than an optical wave-length.
Furthermore the phase imprinting does not produce a density minimum
characteristic to the dark solitons in the region of the phase change. Hence
completely dark stationary solitons are difficult to achieve, which
consequently results in so called grey moving solitons with a shallow density
dip.

It is of a significant interest to be able to create slowly moving, or even
completely stationary solitons in order to test for instance their scattering
properties.  The shapes of the colliding solitons are to be preserved. In
addition, a relative spatial shift is expected. This spatial shift, however, can
only be detected for extremely slow solitons due to the inherent logarithmic
dependence of the spatial shift on the relative velocity between the solitons
\cite{zakharov73,burger02}. The standard phase imprinting also inevitably
creates phonons in the trapped cloud because the constructed initial state is
not the exact soliton solution largely due to the missing density notch
\cite{burger99,denschlag00}.

In this paper we show how states which have the required phase slip and density
profile for solitons can be created by sweeping three laser beams through an
elongated BEC as shown in Fig.~\ref{fig1}. If one of  the beams is in the first
order (TEM10) Hermite-Gaussian mode, its amplitude has a transversal $\pi$
phase slip which will be transferred to atoms thus producing a soliton. More
importantly, with a sequence of three laser beams it is possible to circumvent
the restriction set by the diffraction limit. The laser fields reshape an
atomic wave-function so that it acquires a zero-point.  This leads to a hole in
the atomic density, the width of which is only limited by the intensity ratios
between the incident laser beams due to the geometric nature \cite{ruseckas05}
of the process. The formation of the hole is accompanied by a step-like
(infinitely sharp) $\pi$ phase slip in the atomic wave-function when crossing
the zero-point. The method is particularly useful for creating multicomponent
(vector) solitons of the dark-bright form as well as the dark-dark combination.
In addition it is possible to create in a controllable way two or more slowly
moving dark solitons close to each other for studying their collisional
properties.

\section{Formulation}
\subsection{Outline of the proposed setup}

\begin{figure}
\centering
\includegraphics[angle=270, width=0.70\textwidth]{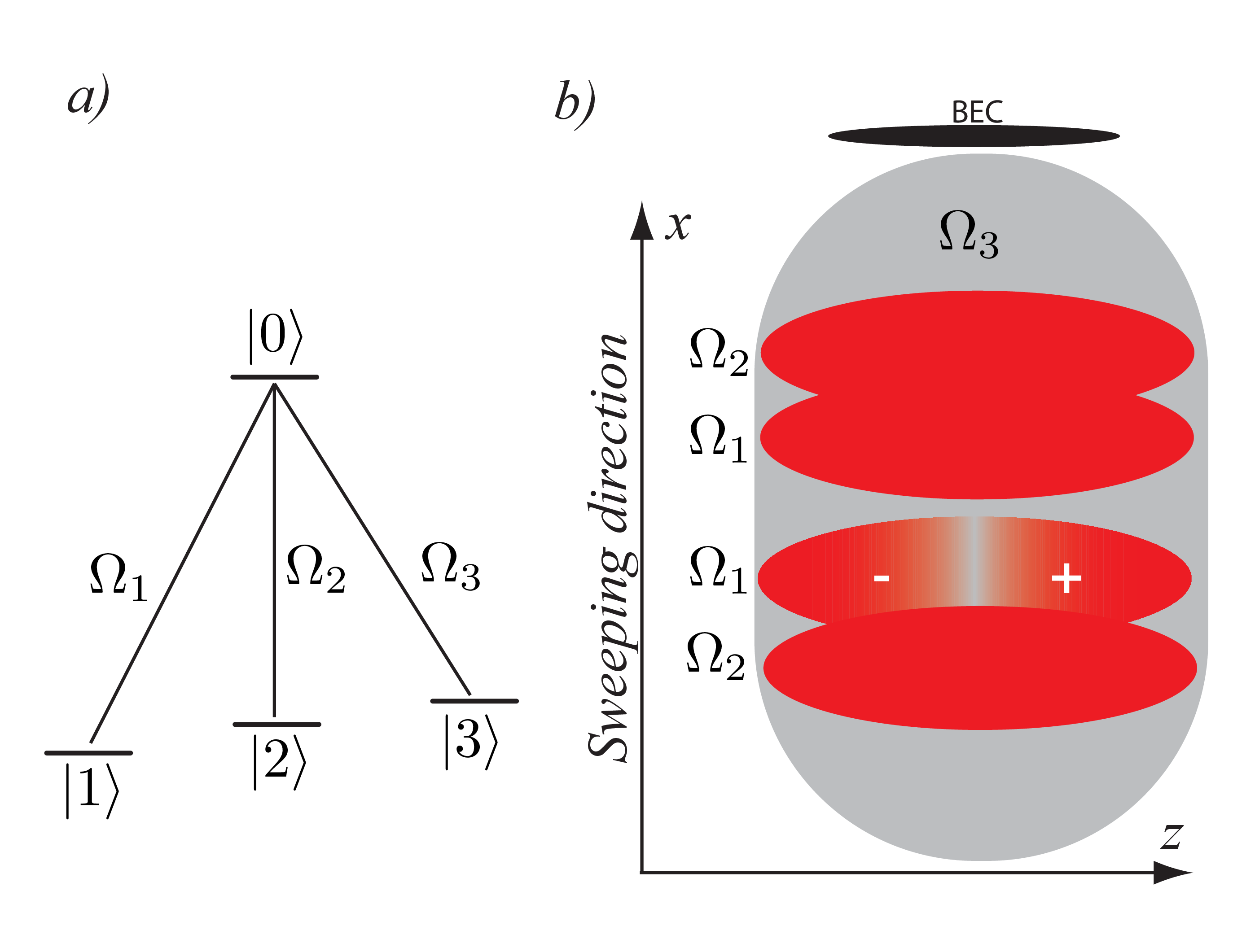}
\caption{a) The level scheme for the three laser beams $\Omega_i$ ($i=1,2,3$).
b) The sequence of laser beams being swept through the BEC involves a
preparation stage $\Omega_2\rightarrow\Omega_1$ and a final stage
$\Omega_1\rightarrow\Omega_2$ which engineers the phase and density of the BEC
to produce a soliton.}
\label{fig1}
\end{figure}

Consider a cigar-shape atomic BEC elongated in the $z$-direction.  To create
solitons in the BEC, we propose to sweep three incident laser beams across the
condensate. The laser beams interact with the condensate atoms in a tripod
configuration \cite{unanyan98,ruseckas05}, i.e.\ the atoms are characterized by
three ground states $|1\rangle$, $|2\rangle$, $|3\rangle$ and an excited state
$|0\rangle$. The $j$-th laser drives resonantly the atomic transition between
the ground state $|j\rangle$ and the excited state $|0\rangle$, see
Fig.~\ref{fig1}a. Initially the atoms forming the BEC are prepared in the
hyperfine ground state $|1\rangle$. Subsequently the lasers are swept through
the BEC in the $x$-direction, i.e.\ perpendicular to the longitudinal axis $z$
of the condensate.

The sweeping process is made of two stages depicted in Fig.~\ref{fig1}b. In the
first stage the lasers $1$ and $2$ are applied in a counter-intuitive sequence
to transfer adiabatically the atoms from the ground state $|1\rangle$ to
another ground state $|2\rangle$. If an additional laser $3$ is on during the
first stage, a partial transfer of atoms from the ground states $|1\rangle$ to
$|2\rangle$ is possible \cite{unanyan98}.  In that case a coherent superposition
of states $|1\rangle$ and $|2\rangle$ is created after completing the first
stage. In the second stage, the lasers $1$, $2$ and $3$ are applied once again
to transfer atoms from the state $|2\rangle$ back to the state $|1\rangle$ and
from the state $|1\rangle$ to the state $|2\rangle$. If the amplitude of one of
these lasers $\Omega_{1}$ or $\Omega_{2}$ changes the sign at $z=z_{0}$, the
BEC picks up a $\pi$ phase shift at this point after the sweeping, and a
soliton can be formed. This is the case e.g.\ if one of the beams is the first
order Hermite-Gaussian beam centered at $z=0$.

It is important to realize that at least two laser fields are needed to
complete the adiabatic transfer of population between the ground states.
Therefore the adiabaticity can be violated in the vicinity of the point
$z=z_{0}$ where one of the Rabi frequencies $\Omega_{1}$ or $\Omega_{2}$ goes
to zero.  Inclusion of the third (support) laser $3$ helps to avoid such a
violation of the adiabaticity.  In fact the atoms would experience absorption
in the vicinity of $z=z_{0}$ if the support laser $3$ was missing during the
second stage.

It  should be mentioned that there are similar previous proposals for creating
vortices in a BEC  via the two-laser Raman processes involving the transfer of
an optical vortex to the atoms \cite{marzlin97,nandi04,andersen06}.  In these
schemes the lasers are far detuned from the single-photon resonance to avoid
the absorption at the vortex core. In our scheme the lasers are in an exact
single-photon resonance, so the use of the third (support) laser is essential
to avoid the losses. An advantage of the resonant scheme is that an efficient
and complete population transfer is possible between the hyperfine ground
states, whereas in the non-resonant case only a fraction of population can be
transferred \cite{andersen06}.

\subsection{Hamiltonian for a tripod atom}

Let us now provide a quantitative description of our scheme.  The $j$-th laser
beam is characterised by the complex Rabi frequency
$\tilde{\Omega}_{j}=\Omega_{j}\exp(i\mathbf{k}_{j}\cdot\mathbf{r}+iS_{j})$ ,
with $j=1,2,3$, where $\Omega_{j}$ is the real amplitude, the phase being
comprised of the local phase $\mathbf{k}_{j}\cdot\mathbf{r}$ as well as the
global (distance-independent) phase $S_{j}$.  In what follows, the Rabi
frequencies $\Omega_{2}$ and $\Omega_{3}$ are considered to be positive:
$\Omega_{2}>0$, $\Omega_{3}>0$. Yet, the Rabi frequency $\Omega_{1}$ is allowed
to be negative. This makes it possible to include an additional $\pi$ phase
shift in the spatial profile of the first beam when crossing the zero-point at
$z=z_{0}$.

The electronic Hamiltonian of a tripod atom reads in the interaction
representation:
\begin{equation}
\hat{H}_{e}=-\hbar(\tilde{\Omega}_{1}|0\rangle\langle1|
+\tilde{\Omega}_{2}|0\rangle\langle2|+\tilde{\Omega}_{3}|0\rangle\langle3|)
+\mathrm{H.c.}
\label{H}
\end{equation}
The tripod atoms have two degenerate dark states $|D_{1}\rangle$ and
$|D_{2}\rangle$ of zero eigen-energy ($\hat{H}_{e}|D_{n}\rangle=0$) containing
no excited-state contribution \cite{unanyan98,ruseckas05},
\begin{eqnarray}
|D_{1}\rangle & = & \frac{1}{\sqrt{1+\zeta^{2}}}
\left(|1\rangle^{\prime}-\zeta|2\rangle^{\prime}\right)\label{D1}\\
|D_{2}\rangle & = & \frac{1}{\sqrt{1+\zeta^{2}}}\left(\xi_{3}
\left(\zeta|1\rangle^{\prime}+|2\rangle^{\prime}\right)
-\xi_{2}(1+\zeta^{2})|3\rangle^{\prime}\right)\,,
\label{D2}
\end{eqnarray}
where $|j\rangle^{\prime}=|j\rangle\exp(i(\mathbf{k}_{3}-\mathbf{k}_{j})
\cdot\mathbf{r}+i(S_{3}-S_{j}))$ (with $j=1,2,3$) are the modified atomic
state-vectors accommodating the phases of the incident laser fields,
$\zeta=\Omega_{1}/\Omega_{2}$ is the ratio between the Rabi frequencies of the
first and second fields, and $\xi_{j}$ are the normalised Rabi frequencies
($j=1,2,3$),
\begin{equation}
\xi_{j}=\frac{\Omega_{j}}{\Omega},\qquad
\Omega=\sqrt{\Omega_{1}^{2}+\Omega_{2}^{2}+\Omega_{3}^{2}}
\label{ksi-j}
\end{equation}
with $\xi_{3}>0$ and $-\infty<\zeta<+\infty$. The atomic dark states
$|D_{1}\rangle$ and $|D_{2}\rangle$ depend on the centre of mass coordinate
$\mathbf{r}$ through the spatial dependence of the Rabi frequencies
$\Omega_{j}$ and state-vectors $|j\rangle^{\prime}$.

\subsection{General equations of motion}

The full atomic state-vector of a multicomponent BEC is
$|\Phi(\mathbf{r},t)\rangle=\sum_{j=1}^{4}|j\rangle\Psi_{j}(\mathbf{r},t)$,
where the constituent wave functions $\Psi_{j}(\mathbf{r},t)$ describe the
translational motion of the BEC in the internal state $|j\rangle$ of the tripod
scheme. The wave functions $\Psi_{j}(\mathbf{r},t)$ obey a multicomponent
Gross-Pitaevski equation of the form
\begin{equation}
i\hbar\frac{\partial}{\partial t}|\Phi(\mathbf{r},t)\rangle=
\left[\frac{1}{2M}\nabla^{2}+\hat{H}_{e}
+\hat{V}\right]|\Phi(\mathbf{r},t)\rangle ,
\label{eq:GP-Eq}
\end{equation}
where $\hat{H}_{e}$ from Eq.~(\ref{H}) describes the light-induced transitions
between the different internal states of atoms. The diagonal operator
\begin{equation}
\hat{V}=\sum_{l>j=0}^{3}(V_{j}+g_{jl}|\Psi_{l}|^{2})|j\rangle\langle
j|\,.
\label{eq:V}
\end{equation}
accommodates the trapping potential $V_{j}(\mathbf{r})$ for the $j$-th internal
state, as well as the nonlinear interaction between the components $j$ and $l$
characterised by the strength $g_{jl}=4\pi\hbar^{2}a_{jl}/m$, with $a_{jl}$
being the corresponding scattering length.

\section{Time-evolution of the atom-light system}
\subsection{Adiabatic approximation for the dark states}

We shall apply the adiabatic approximation \cite{ruseckas05,juzeliunas05pra,juzeliunas05ljp}
under which atoms evolve within
their dark-state manifold during the sweeping. This is legitimate if the total
Rabi frequency $\Omega$ is sufficiently large compared to the inverse sweeping
duration $\tau^{-1}_{\mathrm{sweep}}$. The full atomic state-vector can then be
expanded as: $|\Phi(\mathbf{r},t)\rangle=\sum_{n=1}^{2}
\Psi_{n}^{(D)}(\mathbf{r},t)|D_{n}(\mathbf{r},t)\rangle$, where a composite
wavefunction $\Psi_{n}^{(D)}(\mathbf{r})$ describes the translational motion of
an atom in the dark state $|D_{n}(\mathbf{r},t)\rangle$.  The atomic centre of
mass motion is thus represented by a two-component wave-function
\begin{equation}
\Psi=\left(\begin{array}{c}
\Psi_{1}^{(D)}\\
\Psi_{2}^{(D)
}\end{array}\right)\,.
\label{multicomponent}
\end{equation}
obeying the following equation of motion \cite{ruseckas05}:
\begin{equation}
i\hbar\frac{\partial}{\partial t}\Psi=
\left[\frac{1}{2M}(-i\hbar\nabla-\mathbf{A})^{2}
+V(\mathbf{r})+\phi-\beta\right]\Psi\,,
\label{Eq-dark}
\end{equation}
where the effective vector potential $\mathbf{A}$ and the matrix $\beta$ are
the $2\times2$ matrices appearing due to the spatial and temporal dependence of
the dark states: $\mathbf{A}_{n,m}=i\hbar\langle D_{n}(\mathbf{r},t)|\nabla
D_{m}(\mathbf{r},t)\rangle$ and $\beta_{n,m}=i\hbar\langle
D_{n}(\mathbf{r},t)|\partial/\partial t D_{m}(\mathbf{r},t)\rangle$.  The
former $\mathbf{A}$ is known as the Mead-Berry connection
\cite{berry84,mead91}, whereas the latter matrix $\beta$ is responsible for the
geometric phase \cite{wilczek84}. The $2\times2$ matrix $\phi$ is the effective
trapping potential (explicitly presented in Ref.~\cite{ruseckas05}) appearing
due to the spatial dependence of the dark states. Assuming that all three beams
co-propagate ($\mathbf{k}_1 \approx \mathbf{k}_2 \approx \mathbf{k}_3 $), the
effective vector potential \cite{ruseckas05} reduces to
\begin{equation}
\mathbf{A}=\hbar\frac{\xi_{3}}{1+\zeta^{2}}\nabla\zeta\left(
\begin{array}{cc}
0 & i\\
-i & 0
\end{array}
\right)\,
\label{A-tripod}
\end{equation}
Lastly, the $2\times2$ matrix $V$ originating from the operator $\hat{V}$,
Eq.~(\ref{eq:V}), accommodates the trapping potential for the dark states
\cite{ruseckas05} as well as the atom-atom coupling.

\subsection{Time-evolution during the sweeping}

Suppose the incident laser beams are swept through a trapped BEC along the $x$
axis with a velocity $\mathbf{v}$, as shown in Fig.~\ref{fig1}b. This can be
done either by shifting in the transversal ($x$) direction the laser beams
propagating along the $y$ axis or by applying a set of laser pulses of the
appropriate shape and sequence propagating in the $x$ direction. In the latter
case, the sweeping velocity $v$ will coincide with the speed of light. In both
cases the adiabatic dark states depend on time in the following way:
$|D_{n}(\mathbf{r},t)\rangle \equiv|D_{n}(\mathbf{r}')\rangle$, where
$\mathbf{r}'=(x',y,z)\equiv(x-vt,y,z)$ is the atomic coordinate in the frame of
the moving laser fields.  Let us assume that the time,
$\tau_{\mathrm{sweep}}=d/v$, it takes to sweep the laser beams through a BEC of
the width $d$, is small compared to the time associated with the BEC chemical
potential $\tau_{\mu}=\hbar/\mu$ which is typically of the order of
$10^{-5}\,\mathrm{s}$. In that case one can neglect the dynamics of the atomic
centre of mass during the sweeping. Consequently the time evolution of the
multicomponent wave-function during the sweeping is governed by the matrix-term
$\beta=-vA_{x}$ featured in Eq.~(\ref{Eq-dark}), giving
\begin{equation}
i\hbar\partial_t\Psi=vA_{x}\Psi\,,
\label{eq:fast}
\end{equation}
where the $A_{x}$ the effective vector potential along the sweeping direction.

In passing we note that the subsequent time evolution of the BEC after the
two-stage sweeping will be described by the general Gross-Pitaevski equation
(\ref{eq:GP-Eq}) with the light fields off ($\hat{H}_{e}=0$), as we shall do in
Section~\ref{sect4}.

Returning to Eq.~(\ref{eq:fast}), since $vA_{x}$ commutes with itself at
different times, one can relate the wave-function $\Psi(t)$ at a final time
$t=t_{f}$ to the one at the initial time $t=t_{i}$ as
\begin{equation}
\Psi(\mathbf{r},t_{f})=\exp\left(-i\Theta\right)\Psi(\mathbf{r},t_{i})\,,
\label{solution-formal1}
\end{equation}
where the exponent $\Theta$ is a $2\times2$ Hermitian matrix
\begin{equation}
\Theta=\frac{1}{\hbar}\int_{t_{i}}^{t_{f}}A_{x}(\mathbf{r}-\mathbf{v}t)vdt
=\frac{1}{\hbar}\int_{x_{f}}^{x_{i}}A_{x}(\mathbf{r'})
dx'\,.
\label{Q-A}
\end{equation}
and the integration is over the sweeping path $\mathbf{r}'=(x-vt,y,z)$ from
$x_{f}=x-vt_{f}$ to $x_{i}=x-vt_{i}$. In most cases of interest the initial and
final times can be considered to be sufficiently remote, so that the spatial
integration can be from $x_{f}=-\infty$ to $x_{i}=+\infty$.

\subsubsection{The first stage}

Let us now analyze the proposed two-stage setup in more details. In the first
stage both Rabi frequencies $\Omega_{1}$ and $\Omega_{2}$ are positive. The
lasers $1$ and $2$ are applied in a counterintuitive order (see
Fig.~\ref{fig1}b), where the ratio $\zeta=\Omega_{1}/\Omega_{2}$ changes from
$\zeta(t'_{i})=0$ to $\zeta(t'_{f})=+\infty$. On the other hand, the laser $3$
is dominant for both the initial and final times where
$\xi_{3}=\Omega_{3}/\Omega=1$. Initially the BEC has the wave-function
$\Psi(\mathbf{r})$ and is in the internal atomic ground state $|1\rangle$ which
coincides with the first dark state at the initial time $t'_{i}$,
i.e.\ $|D_{1}(\mathbf{r},t'_{i})\rangle=|1\rangle$.  The full initial atomic
state-vector is therefore
$|\Phi(\mathbf{r},t'_{i})\rangle=\Psi(\mathbf{r})|D_{1}(\mathbf{r},t'_{i})\rangle$.
This provides the following initial condition for the multicomponent
wave-function:
\begin{equation}
\Psi(\mathbf{r},t'_{i})=\left(
\begin{array}{c}
\Psi(\mathbf{r})\\
0
\end{array}\right)\,.
\label{solution-initial}
\end{equation}
Equations (\ref{A-tripod}) and (\ref{solution-formal1})--(\ref{solution-initial})
yield the multicomponent wave-function after the first stage
\begin{equation}
\Psi(\mathbf{r},t'_{f})=\Psi(\mathbf{r})\left(
\begin{array}{c}
\cos\beta\\
-\sin\beta
\end{array}\right)\,,
\label{solution-specific}
\end{equation}
where
\begin{equation}
\beta=\int_{-\infty}^{+\infty}\xi_{3}\frac{\partial\arctan\zeta}{\partial
x'}dx'\,
\label{alpha-z}
\end{equation}
is the mixing angle between the dark states acquired in the first stage.

Suppose we have the following laser beams. The second beam $\Omega_{2}$ is the
Gaussian beam characterised by a waist $\sigma_{z}$ in the $z$ direction. The
beam is centered at $x'=\bar{x}+\Delta$ in the sweeping direction and at $z=0$
in the $z$ direction,
\begin{equation}
\Omega_{2}=Ae^{-z^{2}/\sigma_{z}^{2}-(x'-\bar{x}-\Delta)^{2}/\sigma_{x}^{2}}.
\label{Omega-2}
\end{equation}
The first beam $\Omega_{1}$ is characterised by the same amplitude $A$, the
same waist $\sigma_{z}$ and the same width $\sigma_{x}$. Yet it is centered at
$x'=\bar{x}-\Delta$ in the sweeping direction, where $2\Delta$ is the separation
between the two beams. The beam waists should be of the order of the condensate
length (or larger) in the $z$-direction, so that the whole condensate is
illuminated by the beams.

The third beam is considered to change little along the sweeping direction $x$.
Furthermore it has the same width $\sigma_{z}$ in the $z$-direction as the
first two beams
\begin{equation}
\Omega_{3}=A\kappa e^{-z^{2}/\sigma_{z}^{2}}.
\label{Omega-3}
\end{equation}

\begin{figure}
\centering
\includegraphics[width=0.70\textwidth]{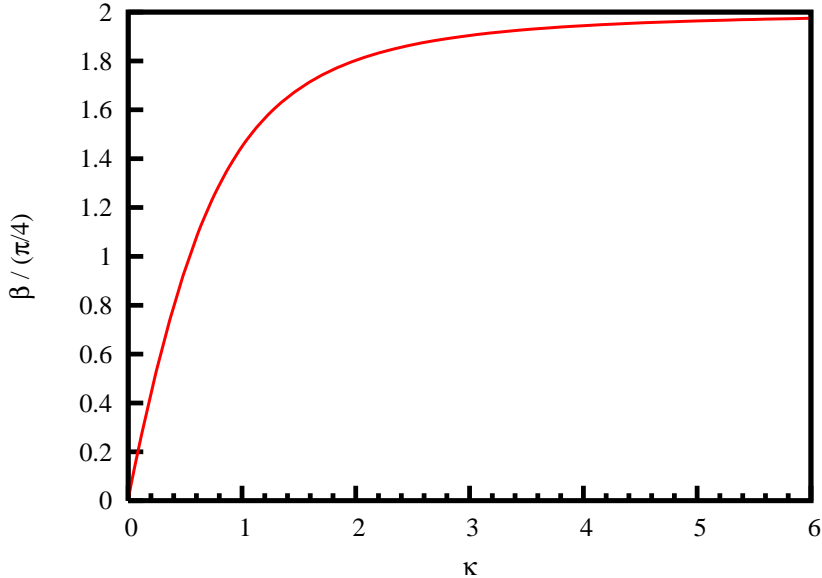}
\caption{Dependence of the mixing angle $\beta$ on the relative amplitude of
the third beam $\kappa$. The spatial separation between the first and the
second beams is taken to be $2\Delta=1.2\sigma_x$ where $\sigma_x$ is the width
of the beams in the sweeping direction.}
\label{fig-beta}
\end{figure}

The first stage is aimed at creating a superposition of states $|1\rangle$ and
$|2\rangle$. Since we take all the beams to be the Gaussian beams characterized
by the same widths $\sigma_{z}$, the Rabi frequency ratios $\Omega_2/\Omega_1$
and $\Omega_3/\Omega_1$ have no $z$-dependence.  As a result the acquired
mixing angle $\beta$ has no $z$-dependence, i.e.\ it is uniform along the BEC.
The magnitude of $\beta$ depends on the relative intensity of the third laser.
If the third laser is weak ($\xi_{3}=\Omega_{3}/\Omega\rightarrow0$ at the
crossing point where $\zeta=\Omega_{1}/\Omega_{2}=1$), the mixing between the
states $|1\rangle$ and $|2\rangle$ is small: $\beta\ll 1$. On the other hand,
if the Rabi frequency $\Omega_{3}$ is comparable with $\Omega_{1}$ and $\Omega_{2}$
at the crossing point where $\zeta=\Omega_{1}/\Omega_{2}=1$, the mixing can be
close to its maximum: $\beta\approx\pi/4$. In this way, one can control the
mixing angle by changing the intensity of the third beam, as one can see from
Fig.~\ref{fig-beta}.

\subsubsection{The second stage}

In the second stage the Rabi frequency $\Omega_{1}$ can be both positive and
negative depending on the transversal coordinate $z$. The laser $1$ is now
applied first, so that the ratio $\zeta=\Omega_{1}/\Omega_{2}$ changes from
$\zeta(t_{i})=\pm\infty$ to $\zeta(t_{f})=0$ in the second stage. Again the
third laser dominates for the initial and final times: $\Omega_{3}/\Omega=1$.
The second stage takes place immediately after completing the first stage, so
the multicomponent wave-function of the first stage (\ref{solution-specific})
serves as an initial condition for the second stage.

Equations (\ref{solution-formal1}), (\ref{Q-A}) and (\ref{solution-specific})
together with (\ref{D1}) and (\ref{D2}) yield the total state vector after the
second stage:
\begin{eqnarray}
|\Phi(\mathbf{r},t_{f})\rangle & = & |1\rangle\Psi(\mathbf{r})
\left(\sin\gamma\cos\beta-e^{i\nu_{12}}\cos\gamma\sin\beta\right)\nonumber\\
&  & -|2\rangle\Psi(\mathbf{r})\left(\cos\gamma\cos\beta
+e^{i\nu_{12}}\sin\gamma\sin\beta\right)\:.
\label{state-vector-final-bare-basis1}
\end{eqnarray}
where $\nu_{12}=S_1-S_2+S_{2}'-S_{1}'$ is the phase mismatch between the Rabi
frequencies $\Omega_{1}$ and $\Omega_2$ in the first and second stages.  The
resulting mixing angle acquired in the second stage is
\begin{equation}
\gamma\equiv \gamma_z=\int_{-\infty}^{+\infty}(1-\xi_{3})
\frac{\partial\arctan\zeta}{\partial x'}dx'\,.
\label{gamma-z}
\end{equation}
If the first and second lasers are weak ($\Omega_{3}/\Omega\rightarrow1$ at the
crossing point where $\zeta=\Omega_{1}/\Omega_{2}=1$), the mixing angle is
small $\gamma_{z}\ll 1$.  On the other hand, if first and second lasers are
strong at this point, we have $\gamma_{z}\rightarrow\mp\pi/2$.  The change in
sign of $\gamma_{z}$ will introduce a phase shift which is needed to create
solitons.

In the second stage the first beam $\Omega_{1}$ is a first-order (in the $z$
direction) Hermite-Gaussian beam centered at $z=0$ and
$x'=\tilde{x}+\tilde{\Delta}$
\begin{equation}
\Omega_{1}=A\frac{z}{B}e^{-z^{2}/\sigma_{z}^{2}
-(x'-\tilde{x}-\tilde{\Delta})^{2}/\sigma_{x}^{2}},
\label{Omega-1}
\end{equation}
where $z=\pm B$ represents a distance where $\Omega_{1}=\pm\Omega_{2}$ for
$x'=\tilde{x}$. In most cases of interest the distance $B$ is much smaller than
the waist of the beams: $B\ll\sigma_z$.  The second beam $\Omega_{2}$ is the
ordinary Gaussian beam centered at $z=0$ along the BEC and
$x'=\tilde{x}-\tilde{\Delta}$ in the sweeping direction
\begin{equation}
\Omega_{2}=Ae^{-z^{2}/\sigma_{z}^{2}-(x'-\tilde{x}+\tilde{\Delta})^{2}/\sigma_{x}^{2}},
\end{equation}
where $2\tilde{\Delta}$ is the separation between the two beams. The ratio
between the Rabi frequencies reads then
\begin{equation}
\zeta=\frac{\Omega_{1}}{\Omega_{2}}=\frac{z}{B}
e^{4\tilde{\Delta}(x'-\tilde{x})/\sigma_{x}^{2}},
\label{zeta}
\end{equation}
Equation~(\ref{zeta}) provides the following limiting cases:
\begin{equation}
\zeta\equiv\zeta(z,x')=\left\{
\begin{array}{cc}
0, & \mathrm{for}\quad x'\rightarrow+\infty\,,\\
\pm\infty,&  \mathrm{for}\quad x'\rightarrow-\infty\,.
\end{array}\right.
\label{zeta-lim}
\end{equation}
Finally let us determine the crossing point where $\Omega_{1}=\Omega_{2}$.
Using Eq.~(\ref{zeta}), the condition $|\zeta|=1$ yields the crossing point
$x'=x'_{\mathrm{cr}}$ for a fixed $z$ coordinate:
\begin{equation}
x'_{\mathrm{cr}}=\tilde{x}+\frac{\sigma^{2}}{4\tilde{\Delta}}\ln\frac{z}{B}\,.
\label{y-cr}
\end{equation}
Specifically, if $z=B$, the crossing point is: $x'_{\mathrm{cr}}=\tilde{x}$.
Since $B\ll\sigma$, the Rabi frequencies at $z=B$ and $x'=\tilde{x}$ are:
\begin{equation}
\Omega_{1}=\Omega_{2}\approx Ae^{-\tilde{\Delta}^{2}/\sigma_{x}^{2}}.
\label{Omega-1-2-crossing}
\end{equation}

In the next subsection we shall analyse in more detail the multicomponent
wave-function after completing the second stage.

\subsection{Multicomponent wave-function alter the sweeping}

\begin{figure}
\centering
\includegraphics[width=0.70\textwidth]{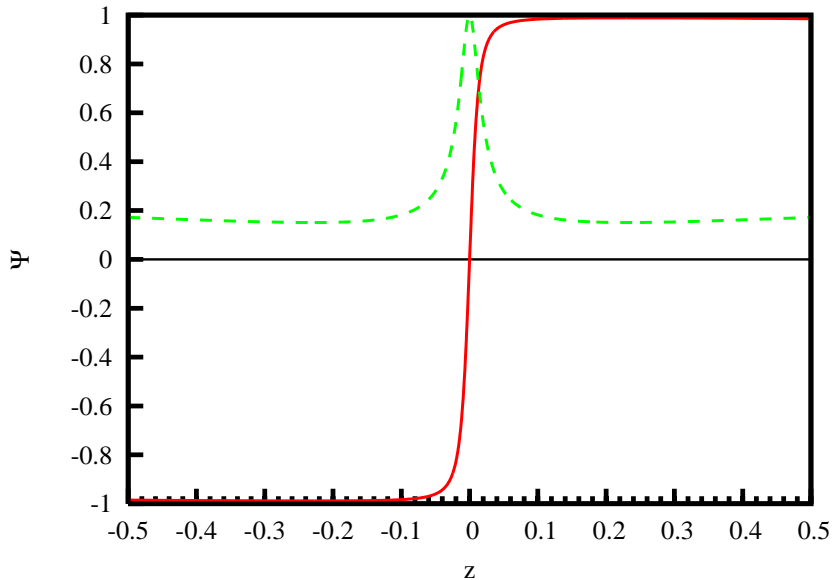}
\caption{Multicomponent wave-function after completing the second stage in the
case where the second component is populated after the first stage ($\beta=0$)
and there is no phase mismatch between the lasers of the first and second
stages ($\nu_{12}=0$). The second and third laser beams are taken to be the
Gaussian beams with equal widths $\sigma_{z}$. The first laser beam is the
first order Hermite-Gaussian beam with same width $\sigma_{z}$. The parameters
used are $2\tilde{\Delta}/\sigma_x = 1.2$, $B/\sigma_z=0.1$ and $\kappa=0.1$.
The wave function of the first (second) component is plotted in a solid
(dashed) line.}
\label{fig-dark-bright}
\end{figure}

Suppose that there is no phase mismatch between the lasers of the first and
second stages: $\nu_{12}=0$. In that case
Eq.~(\ref{state-vector-final-bare-basis1}) yields
\begin{equation}
|\Phi(\mathbf{r},t_{f})\rangle=
\Psi(\mathbf{r})[-\sin(\gamma_{z}-\beta)|1\rangle
+\cos(\gamma_{z}-\beta)|2\rangle]\,,
\label{state-vector-final-bare-basis2b}
\end{equation}
If $\beta=0$, the second component is populated after the first stage.  After
the whole sweeping the state-vector then takes the form
\begin{equation}
|\Phi(\mathbf{r},t_{f})\rangle=\Psi(\mathbf{r})[-\sin\gamma_{z}|1\rangle
+\cos\gamma_{z}|2\rangle]\,.
\label{state-vector-final--nu=0-beta=pi/2}
\end{equation}
In this case the first component alters the sign at $z=z_{0}$ where the Rabi
frequency $\Omega_{1}$ or $\Omega_{2}$ (and hence $\gamma_{z}$) crosses the
zero-point. On the other hand, the second component is maximum at this point
and symmetrically decays to zero away from this point. Such a multicomponent
wave-function has a shape close to that of a soliton of the dark-bright form
(see Fig.~\ref{fig-dark-bright}).  This will indeed lead to the formation of
such a soliton, as we shall from the analysis of the subsequent time-evolution
presented in the next Section.

\begin{figure}
\centering
\includegraphics[width=0.70\textwidth]{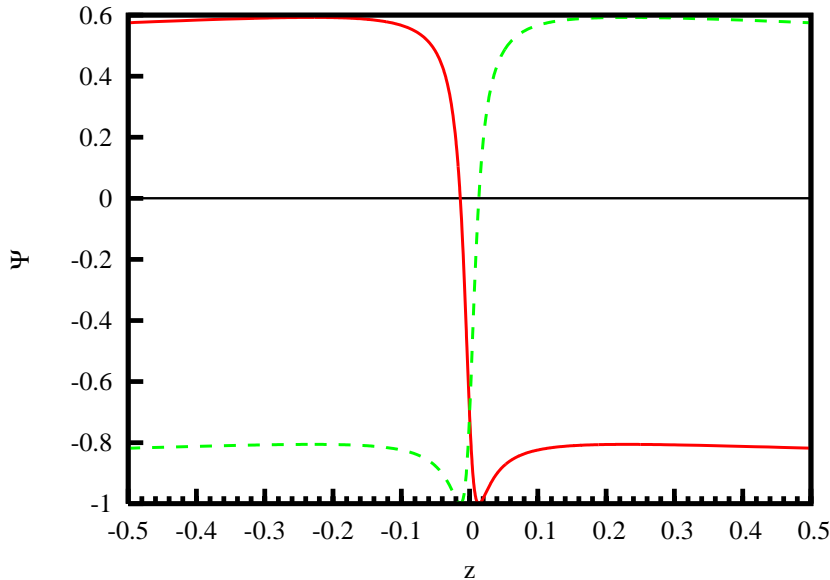}
\caption{Multicomponent wave-function after completing the second stage in the
case where where both components are initially populated after the first stage
($\beta=\pi/4$) and there is no phase mismatch between the lasers of the first
and second stages ($\nu_{12}=0$). The second and third laser beams are taken to
be the Gaussian beams with equal widths $\sigma_{z}$. The first laser beam is
the first order Hermite-Gaussian beam with same width $\sigma_{z}$. The
parameters used are $2\tilde{\Delta}/\sigma_x = 1.2$, $B/\sigma_z=0.1$ and
$\kappa=0.1$. The wave function of the first (second) component is plotted in a
solid (dashed) line.}
\label{fig-dark-dark}
\end{figure}

On the other hand, $\beta=\pi/4$ corresponds to the case where both components
are initially populated with equal probabilities. Thus we have after the
sweeping:
\begin{equation}
|\Phi(\mathbf{r},t_{f})\rangle=
-\Psi(\mathbf{r})[-\sin(\gamma_{z}-\pi/4)|1\rangle
+\sin(\gamma_{z}+\pi/4)|2\rangle]\,.
\label{state-vector-final-bare-basis3}
\end{equation}
In that case \emph{both components} of the wave-function acquire a $\pi$
\emph{phase shift} in a vicinity of $z=z_{0}$ where $\Omega_{1}=0$, as one can
see clearly in the Fig.~\ref{fig-dark-dark} Note that the zero-points of each
component are slightly shifted with respect to each other.  This makes it
possible to produce two component dark-dark solitons oscillating around each
other, as we shall see in the following Section.

If  $\beta=\pi/4$ , yet there is a $\pi/2$ phase mismatch ($\nu_{12}=\pi/2$),
Eq.~(\ref{state-vector-final-bare-basis1}) reduces to
\begin{equation}
|\Phi(\mathbf{r},t_{f})\rangle=
-\Psi(\mathbf{r})e^{i\gamma_{z}}\frac{1}{\sqrt{2}}
\left[|2\rangle-i|1\rangle\right]\,.
\label{state-vector-final-bare-basis2e}
\end{equation}
In that case both components are characterised by the same spatial modulation
$\exp\left(i\gamma_{z}\right)$ and have a relative phase $\pi/2$ after the
sweeping. Therefore both components initially have the same velocity
distribution proportional to $\nabla\gamma_{z}$.  Furthermore there is no hole
in the atomic density of neither component after the sweeping, similar the case
in the phase imprinting techniques.

In this way, the creation of solitons can be controlled by changing the mixing
angle $\beta$ and the phase mismatch $\nu_{12}$

\section{Subsequent dynamics and soliton formation}

\label{sect4}
The optical preparation of the initial state of the two-component Bose-Einstein
condensate described in the previous section, is fast compared to any
characteristic dynamics in the Bose-Einstein condensate.  This is the case if
the time $\tau_{\mathrm{sweep}}=d/v$ it takes to sweep the laser beams through
a BEC of the width $d$, is small compared to the time associated with the BEC
chemical potential $\tau_{\mu}=\hbar/\mu$ which is typically of the order of
$10^{-5}\,\mathrm{s}$.  With the prepared initial state and for sufficiently
low temperatures we can therefore describe the subsequent dynamics using a
two-component Gross-Pitaevskii equation \cite{ohberg01}
\begin{eqnarray}
i\hbar\frac{\partial}{\partial t} \Psi_1&=&[-\frac{\hbar^2}{2m}\nabla^2+V(z)
+g_{11}|\Psi_1|^2+g_{12}|\Psi_2|^2]\Psi_1\label{gp1}\\
i\hbar\frac{\partial}{\partial t} \Psi_2&=&[-\frac{\hbar^2}{2m}\nabla^2+V(z)
+g_{22}|\Psi_2|^2+g_{12}|\Psi_1|^2]\Psi_2.\label{gp2}
\end{eqnarray}
The external potential is here chosen to be quadratic in the $z$-direction,
\begin{equation}
V(z)=\frac{1}{2} m\omega^2 z^2,
\end{equation}
where $\omega$ is the trap frequency and $m$ the atomic mass. The two-body
interactions are described by 
\begin{equation}
g_{ij}=\frac{4\pi\hbar^2a_{ij}}{mS}, \quad i,j=\{1,2\} \label{gij}
\end{equation}
with the scattering lengths $a_{ij}$ which represents the intra and inter
collisional interactions between the atoms in the states $1$ and $2$. In
Eq.~(\ref{gij}) we have introduced the effective cross-section $S$ of the
elongated cloud. Strictly speaking the elongated Bose-Einstein condensate is
three dimensional. If, however, the transversal trapping is sufficiently strong,
the dynamics can be considered effectively one dimensional, as in
Eqs.~(\ref{gp1}) and (\ref{gp2}). This requires that the corresponding
transversal ground state energy is much larger than the chemical potential of
the condensate. We choose the normalisation as $\int dz|\Psi_i(z)|^2=N_i$,
where $N_i$ is the particle number in condensate $i$ $(i=1,2)$.

\begin{figure}
\centering
\includegraphics[width=0.45\textwidth]{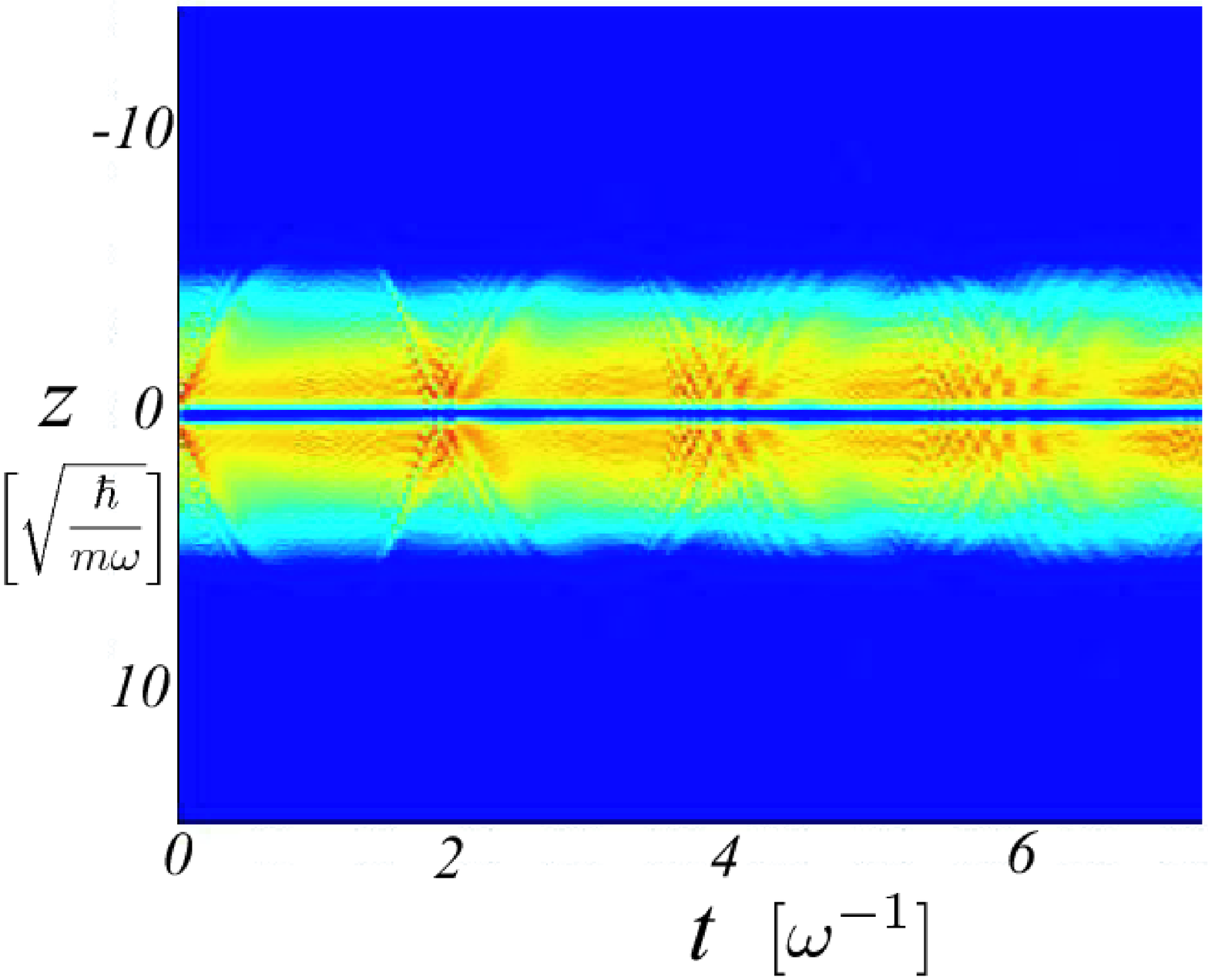}
\includegraphics[width=0.45\textwidth]{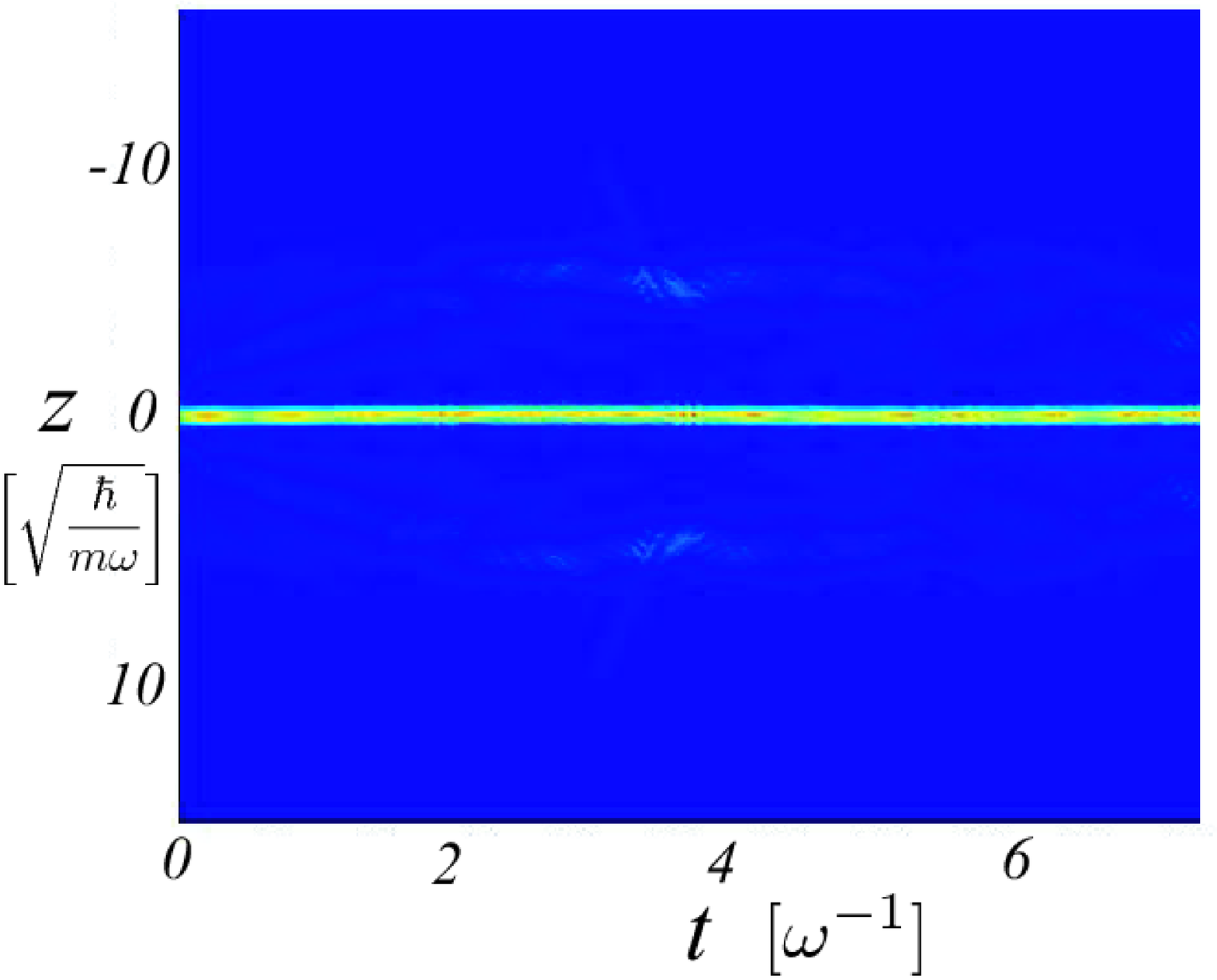}
\caption{The dark-bright soliton. The two figures show the one dimensional
density as a function of time for component $1$ and $2$. The lighter (darker) colours 
correspond to higher (lower) atomic densities.}
\label{beta0}
\end{figure}

\begin{figure}
\centering
\includegraphics[width=0.45\textwidth]{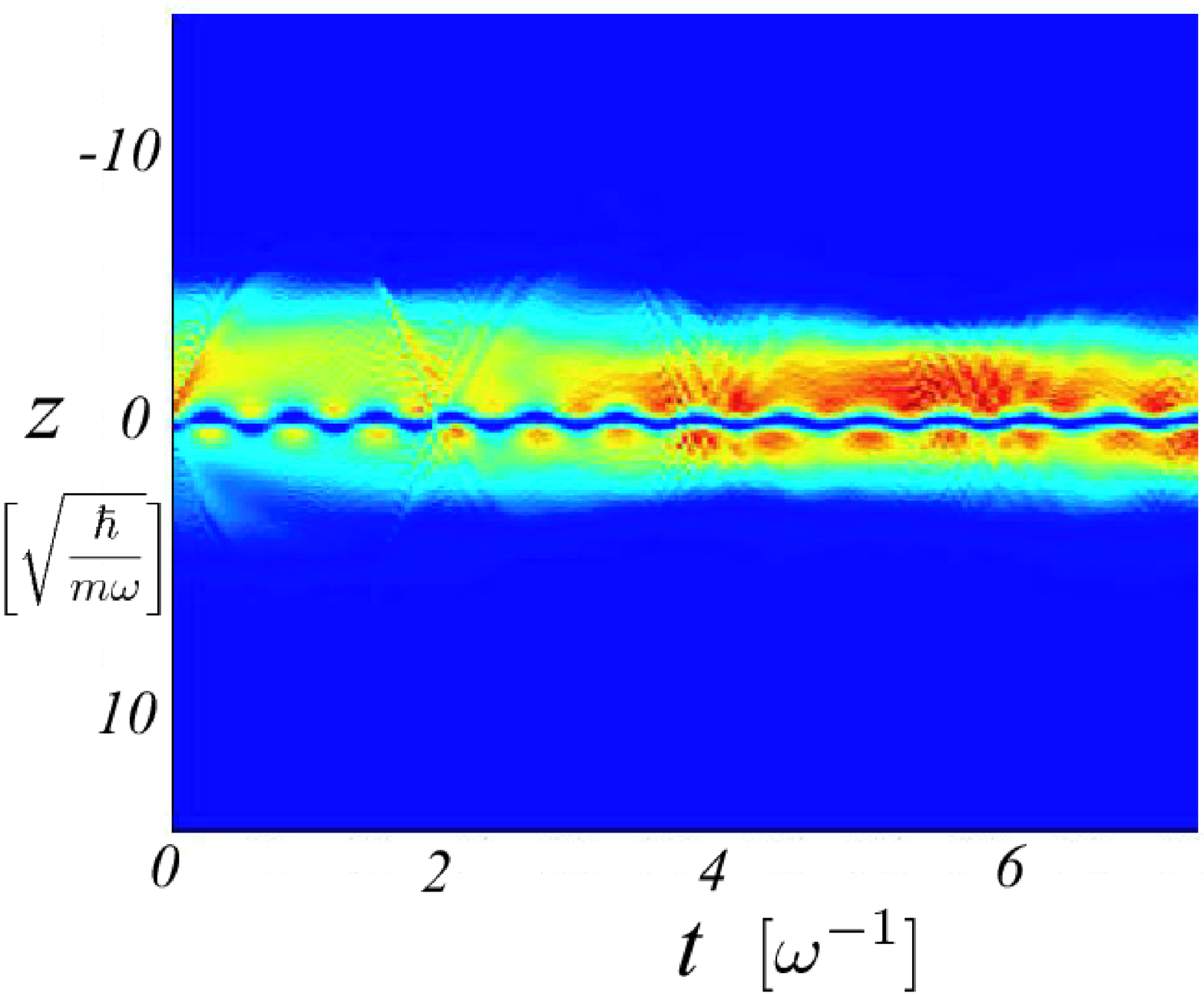}
\includegraphics[width=0.45\textwidth]{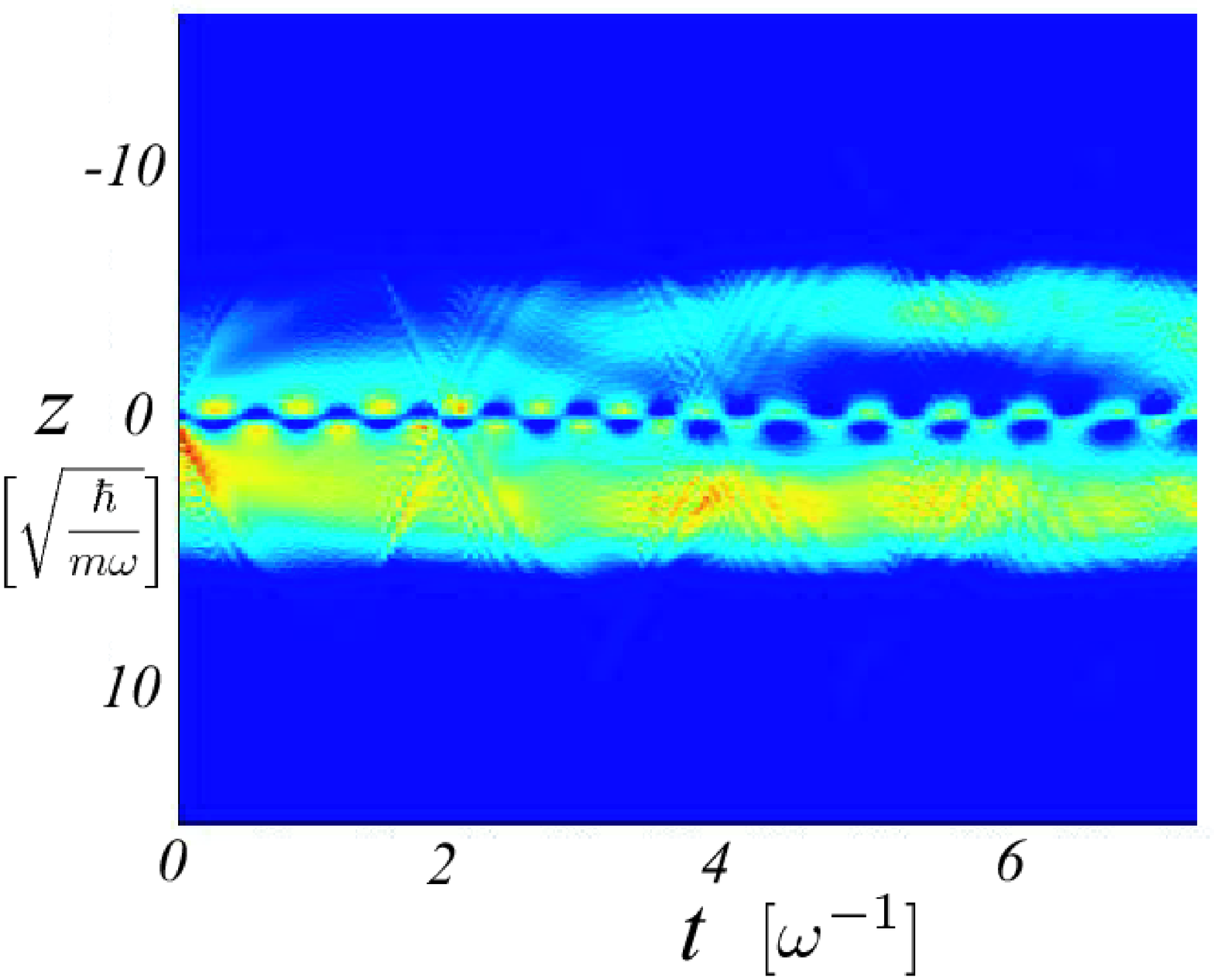}
\caption{The bound state dark-dark soliton. For sufficiently low initial
soliton velocities the two dark solitons perform an oscillatory motion around
each other. The figures show the one dimensional atomic density as a function of time
for component $1$ and $2$. The lighter (darker) colours correspond to higher (lower) densities.}
\label{betapi4nu0}
\end{figure}

\begin{figure}
\centering
\includegraphics[width=0.45\textwidth]{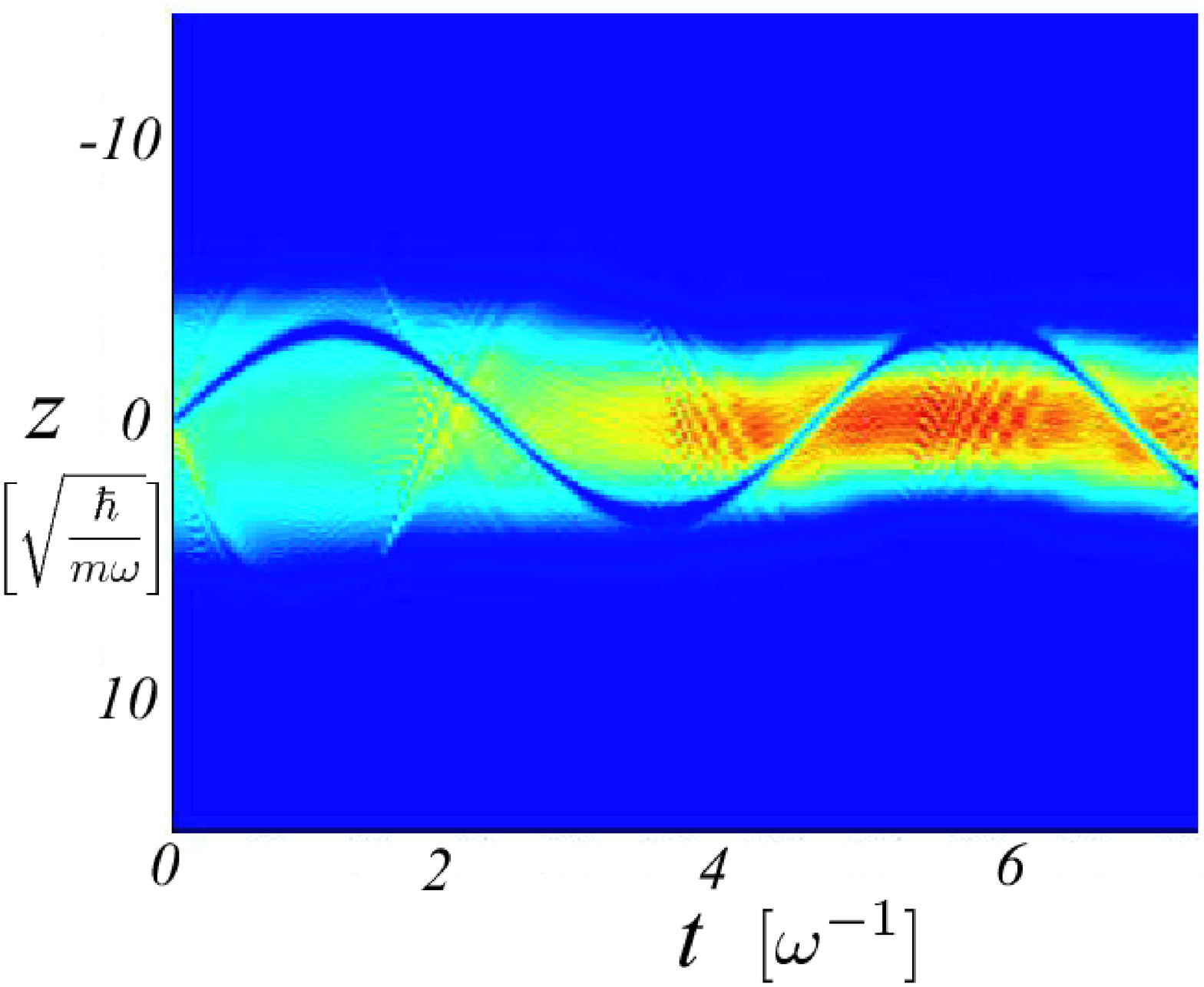}
\includegraphics[width=0.45\textwidth]{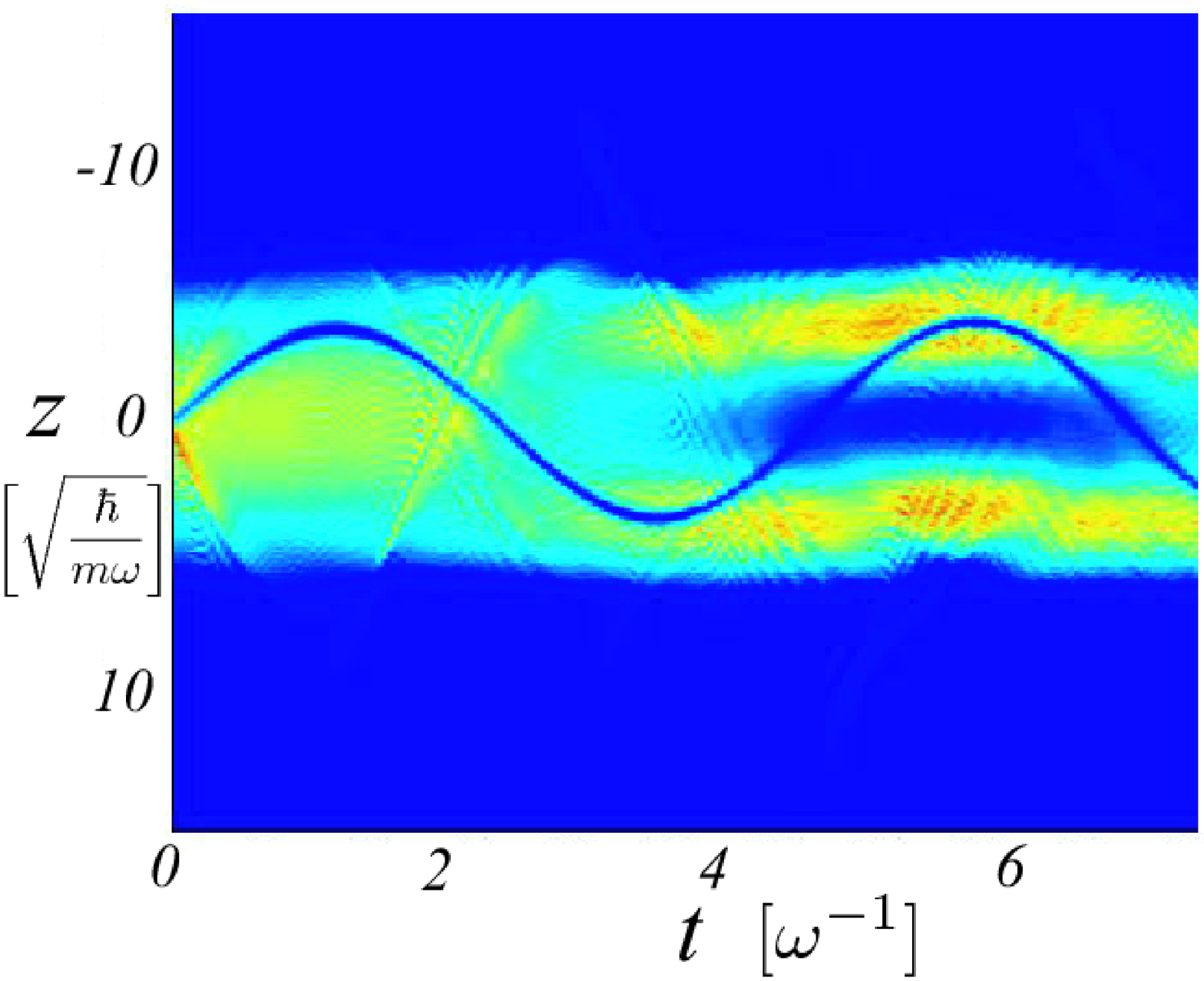}
\caption{The co-propagating dark-dark solitons. If the initial phase gradients
of the two soliton solutions are chosen to be the same the dark solitons
propagate in unison. The two figures show the one dimensional density as a
function of time for component $1$ and $2$. The lighter (darker) colours 
correspond to higher (lower) atomic densities.}
\label{betapi4nupi2}
\end{figure}

With the initial states from the previous section we can simulate the dynamics
of the Bose-Einstein condensate. We consider a condensate with
$g_{11}:g_{12}:g_{22}=1.0:0.97:1.03$ where $g_{12}N_1=286$ and $N_1=N_2$. The
unit of length is $\sqrt{\frac{\hbar}{m\omega}}$ and time is in units of
$\omega^{-1}$. In figure \ref{beta0} we show the dark-bright soliton dynamics
whose initial state is prepared by choosing $\beta=0$ and $\nu_{12}=0$. The
two-component system which has one dark soliton in component $1$ and a bright
soliton in component $2$, is stable, i.e.\ the solitons are stationary. This
shows that the initial state is indeed close to the exact soliton solution. If
the initial state is prepared with $\beta=\pi/4$ and $\nu_{12}=0$, on the other
hand, the dynamics is strikingly different, see figure \ref{betapi4nu0}. In
this case we create two dark solitons with opposite phase gradients, hence
there is an oscillatory motion, sometimes referred to as a soliton molecule.
Such a bound state is only stable if the soliton velocities are low
\cite{ohberg01} which is indeed the case here. Alternatively, with
$\beta=\pi/4$ and $\nu_{12}=\pi/2$, the solitons move in unison as shown in figure
\ref{betapi4nupi2}. The large oscillatory motion appearing in
Fig.~\ref{betapi4nupi2} stems from the fact that the condensate density is not
homogeneous, hence the solitons experience an effective trap \cite{busch00}.

\section{Conclusions}

In summary, we have proposed a new method of creating solitons in elongated
Bose-Einstein Condensates (BECs) by sweeping three laser beams through the BEC.
If one of  the beams is the first order (TEM10) Hermite-Gaussian mode, its
amplitude has a transversal $\pi$ phase slip which will be transferred to the
atoms thus creating a soliton. Using this method it is possible to circumvent
the restriction set by the diffraction limit. The method allows one to create
multicomponent (vector) solitons of the dark-bright form as well as the
dark-dark combination. In addition it is possible to create in a controllable
way two or more slowly moving dark solitons close to each other for studying
the collisional properties. For this the first beam $\Omega_1$ should represent
a superposition of the zero and second order Hermite-Gaussian modes in the
second stage. The soliton collisions will be considered in more details
elsewhere.

\subsection*{Acknowledgements}

This work was supported by the Alexander-von-Humboldt foundation through the
institutional collaborative grant between the University of Kaiserslautern and
the Institute of Theoretical Physics and Astronomy of Vilnius University. 
P.\"O. acknowledges support from the EPSRC and the Royal Society of Edinburgh.

\end{document}